\documentclass[onecolumn,preprint,showpacs,superscriptaddress]{revtex4}

\usepackage[utf8]{inputenc}
\usepackage{mathtools}
\usepackage{graphicx}
\usepackage{caption}
\usepackage{subcaption}
\usepackage{xcolor}
\usepackage{soul}

\begin{document}

\title{Rapidly rotating compact stars in Rastall's gravity}

\author{F. M. da Silva}
\email{franmdasilva@gmail.com}

\affiliation{Departamento de Física, CFM - Universidade Federal de \\ Santa Catarina; C.P. 476, CEP 88.040-900, Florianópolis, SC, Brazil}

\author{L. C. N. Santos}
\email{luis.santos@ufsc.br}

\affiliation{Departamento de Física, CCEN - Universidade Federal da \\ Paraíba; C.P. 5008, CEP  58.051-970, João Pessoa, PB, Brasil.} 

\author{C. C. Barros Jr.}
\email{barros.celso@ufsc.br}

\affiliation{Departamento de Física, CFM - Universidade Federal de \\ Santa Catarina; C.P. 476, CEP 88.040-900, Florianópolis, SC, Brazil} 

\begin{abstract}
In this work we study rapidly rotating stars by considering the Rastall theory of gravity. We obtain and solve the equations by numerical methods for two usual parametrization of polytropic stars. Then the
mass-radius relations, moments of inertia and other results of interest are obtained and compared with the ones for non-rotating stars.
\end{abstract}

\maketitle

\section{Introduction}

Despite the success of General Relativity (GR) that may be considered a fundamental element in the formulation of physics and has been confirmed by many experimental tests, as for example in the recent measurements of gravitational waves by the collaborations Virgo and LIGO (Laser Interferometer Gravitational-Wave Observatory) \cite{abbott2016observation,abbott2019gwtc}, and the even more recent obtaining of the first image of a black hole by the project Event Horizon Telescope \cite{akiyama2019first}, there are still open questions that need to be understood. For instance, GR cannot account for the accelerating expansion of the universe and the rotation curves of galaxies without the introduction of unexplained elements, in this case: dark energy and dark matter, respectively. Due to these seeming limitations, it can be interesting to study if modified theories of gravitation may offer a solution to these questions or at least provide a way to improve our understanding of the theory. 

Because of their very high mass density, compact stars can be highly relativistic objects and thus are good candidates for the study of the effects of modified theories of gravity. There are many studies of modified gravity theories for static stars like in \cite{harada1998neutron,orellana2013structure,momeni2015tolman,oliveira2015neutron,hendi2016modified,singh2019einstein,maurya2020charged,mota2019combined}, just to mention a few, but the studies of this theories on slowly \cite{damour1996tensor,sotani2010slowly,pani2011compact,ali2011slowly,staykov2014slowly,silva2015slowly} and rapidly rotating \cite{doneva2013rapidly,doneva2015iq,yazadjiev2015rapidly,kleihaus2016rapidly,doneva2016rapidly,doneva2018differentially,astashenok2020rotating} stars are still few. From the astrophysical point of view it is important to study rotating stars, since almost all the precise mass measurements of neutron stars that we have nowadays comes from rotating pulsars 
 in binary systems \cite{ozel2016masses}, that can rotate up to $716 Hz$ \cite{hessels2006radio}, in addition, it have been theorized that magnetars can born rapidly rotating \cite{metzger2011protomagnetar,giacomazzo2013formation}. Furthermore, the recent observation of gravitational waves from a compact binary coalescence involving a compact object with a mass within the mass gap of $2.5-5 M_\odot$ \cite{abbott2020gw190814}, led to investigations about the role that rotation and modified theories of gravity combined with a variety of equations of state could play in this observation \cite{tsokaros2020gw190814,nunes2020weighing,astashenok2020extended,clifton2020observational,dexheimer2020gw190814,
koliogiannis2020thermodynamical,demircik2020rapidly,sedrakian2020confronting,astashenok2020extended}.

On the other hand, there are many important studies on rotating stars, as for example in \cite{komatsu1989rapidly,butterworth1976structure} where an useful method of solution of the equations is proposed and many results have been obtained for the considered stellar models, that are essentially based on polytropic equations of state.

In recent years, the theory of gravity proposed by Peter Rastall \cite{rastall1972generalization}, has gained growing interest, with studies of the implications of this theory in the context of black holes \cite{heydarzade2017black,heydarzade2017blackB,ma2017noncommutative,kumar2018rotating,xu2018kerr,ali2020gravitational}, thermodynamic of black holes \cite{bamba2018thermodynamics,lobo2018thermodynamics,soroushfar2019thermodynamic}, wormholes \cite{moradpour2017traversable,halder2019wormhole}, for cosmological scenarios \cite{batista2012rastall,fabris2012rastall,batista2013observational,moradpour2016thermodynamics} as well as some proposals of generalizations of the Rastall's gravity \cite{moradpour2017generalization,lin2020cosmic} and of combining this theory with others modifications of GR \cite{wolf1986non,carames2014brans,mota2019combined} and in particular in the study of compact objects \cite{oliveira2015neutron,mota2019combined,hansraj2019impact,mota2019anisotropic,abbas2018new,salako2018anisotropic,abbas2018isotropic,abbas2019models}, but we still lack a study of the effects of Rastall's theory on rotating stars. 

So, in this work, we will consider these elements in order to formulate a model for rotating stars in the Rastall theory. This paper has the following contents: In Sec. II the formulation of the equations for rotating stars in the GR is done, in Sec. III a brief review of the Rastall theory is shown and in Sec. IV, the formulation of the equations for rotating stars considering the Rastall theory. In Sec. V some aspects of the numerical method are discussed, in Sec. VI the results are presented and in Sec. VII we draft our conclusions.  

\section{Rotating stars in General Relativity}

In this section we are going to present the equations that describe a rapidly rotating star in the framework of the GR and also 
how to solve them by considering the method developed by Komatsu, Eriguchi and Hachisu (KEH) \cite{komatsu1989rapidly}.

In order to describe the space-time of a rotating star in equilibrium it is possible to consider a stationary axially symmetric metric $g_{\mu\nu}$ such as the one presented in \cite{komatsu1989rapidly,stephani2009exact,friedman2013rotating}. The line element can be written in terms of spherical coordinates $\left(t,r,\theta,\phi\right)$, as follows:
\begin{equation}
ds^{2}=-e^{2\nu }dt^{2}+e^{2\alpha }\left( dr^{2}+r^{2}d\theta ^{2}\right)
+e^{2\beta }r^{2}\sin ^{2}\theta \left( d\phi -\omega dt\right) ^{2},
\label{rs1}
\end{equation}
where $\alpha$, $\beta$, $\nu$ and $\omega$ are the metric potentials (or metric functions), which will be determined by solving the Einstein's field equations, and which in the proposed formulation depend only on $r$ and $\theta$. The potential $\omega$ represents the dragging of local inertial frames, also called Lense-Thirring effect \cite{pfister2007history,ciufolini2004confirmation}. All equations in this work are going to be presented in geometrized units, that is, $c=G=1$.

The matter in the stellar interior will be supposed to be a perfect fluid that can be represented by the energy-momentum tensor
\begin{equation}
T^{\mu\nu}=pg^{\mu\nu}+\left( \varepsilon +p\right) U^{\mu}U^{\nu},  \label{rs2}
\end{equation}
where $\epsilon$ and $p$ are the energy density and the pressure, respectively, $g^{\mu\nu}$ is the metric tensor relative to equation $(\ref{rs1})$ and $U^{\mu}$ is the four-velocity of the fluid that has the form \cite{komatsu1989rapidly}
\begin{equation}
U^{\mu}=\frac{dx^{\mu}}{d\tau }=\frac{e^{-\nu }}{\sqrt{1-v^{2}}}\left(
1,0,0,\Omega \right), \label{rs3}
\end{equation}
where $\Omega$ is the angular velocity of an element of mass of the star with respect to a static observer at infinity, and $v$ is the 3-velocity in the referential frame of an observer locally without rotation, usually called ZAMO (zero momentum angular observer), and is given by
\begin{equation}
v=\left( \Omega -\omega \right) r\sin \theta e^{\beta -\nu }.   \label{rs4}
\end{equation}

By considering the line element in equation $(\ref{rs1})$, the energy-momentum tensor in equation $(\ref{rs2})$, and the usual Einstein field equations of GR 
\begin{equation}
    R_{\mu\nu}-\frac{1}{2}g_{\mu\nu}R=8\pi T_{\mu\nu} \label{rs5},
\end{equation}
 we obtain, after some algebra, the following expressions \cite{komatsu1989rapidly}
\begin{equation}
\nabla^{2} \left( \rho e^{\gamma /2}\right) =S_{\rho }\left( r,\mu \right),
\label{rs6}
\end{equation}

\begin{equation}
\left( \nabla^{2} +\frac{1}{r}\frac{\partial }{\partial r}-\frac{1}{r^{2}}\mu
\frac{\partial }{\partial \mu }\right) \gamma e^{\gamma /2}=S_{\gamma
}\left( r,\mu \right),  \label{rs7}
\end{equation}
and

\begin{equation}
\left( \nabla^{2} +\frac{2}{r}\frac{\partial }{\partial r}-\frac{2}{r^{2}}\mu
\frac{\partial }{\partial \mu }\right) \omega e^{\left( \gamma -2\rho
\right) /2}=S_{\omega }\left( r,\mu \right),  \label{rs8}
\end{equation}
where we considered the definitions
\begin{equation}
    \mu=\cos\theta, \label{rs9}
\end{equation}
\begin{equation}
\gamma =\beta +\nu ,  \label{rs11}
\end{equation}
and 

\begin{equation}
\rho =\nu -\beta,  \label{rs12}
\end{equation}
then we have

\begin{equation}
\nabla^{2} =\frac{\partial ^{2}}{\partial r^{2}}+\frac{2}{r}\frac{\partial }{%
\partial r}+\frac{1}{r^{2}}\frac{\partial}{\partial \mu}\left((1-\mu^{2})\frac{\partial}{\partial\mu}\right)+\frac{1}{r^{2}(1-\mu^{2})}\frac{\partial ^{2}}{\partial \phi ^{2}},  \label{rs10}
\end{equation}

\begin{equation}
\begin{split}
S_{\rho }(r,\mu )=& \left[ e^{\gamma /2}8\pi e^{2\alpha }\left( \varepsilon
+p\right) \frac{1+v^{2}}{1-v^{2}}\right.  \\
& +r^{2}(1-\mu ^{2})e^{-2\rho }\left[ \omega _{,r}^{2}+\frac{1}{%
r^{2}}\left( 1-\mu ^{2}\right) \omega _{,\mu }^{2}\right] +\frac{1}{%
r}\gamma _{,r}-\frac{1}{r^{2}}\mu \gamma _{,\mu } \\
& \left. +\frac{\rho }{2}\left\{ 16\pi e^{2\alpha }p-\gamma _{,r}\left( \frac{1}{2}\gamma _{,r}+\frac{1}{r}\right) -\frac{1}{r^{2}%
}\gamma _{,\mu }\left[ \frac{1}{2}\gamma _{,\mu }\left(
1-\mu ^{2}\right) -\mu \right] \right\} \right],
\end{split}
\label{rs13}
\end{equation}

\begin{equation}
S_{\gamma }\left( r,\mu \right) =e^{\gamma /2}\left\{ 16\pi e^{2\alpha }p+%
\frac{\gamma }{2}\left[ 16\pi e^{2\alpha }p-\frac{1}{2}\gamma _{,r}^{2}-\frac{1}{2r^{2}}\left( 1-\mu ^{2}\right) \frac{1}{2}\gamma
_{,\mu }^{2}\right] \right\},  \label{rs14}
\end{equation}
and

\begin{equation}
\begin{split}
S_{\omega }\left( r,\mu \right) =& e^{\left( \gamma -2\rho \right) /2}\left[
-16\pi e^{2\alpha }\frac{\left( \Omega -\omega \right) \left( \varepsilon
+p\right) }{1-v^{2}}\right.  \\
& +\omega \left\{ -8\pi e^{2\alpha }\frac{\left[ \left( 1+v^{2}\right)
\varepsilon +2v^{2}p\right] }{1-v^{2}}-\frac{1}{r}\left( 2\rho _{,r}+\frac{1}{2}\gamma _{,r}\right) \right.  \\
& +\frac{1}{r^{2}}\mu \left( 2\rho _{,\mu }+\frac{1}{2}\gamma
_{,\mu }\right) +\frac{1}{4}\left( 4\rho _{,r}^{2}-\gamma
_{,r}^{2}\right) +\frac{1}{4r^{2}}\left( 1-\mu ^{2}\right) \left(
4\rho _{,\mu }^{2}-\gamma _{,\mu }^{2}\right)  \\
& \left. \left. -r^{2}\left( 1-\mu ^{2}\right) e^{-2\rho }\left[ \omega
_{,r}^{2}+\frac{1}{r^{2}}\left( 1-\mu ^{2}\right) \omega _{,\mu }^{2}\right] \right\} \right].
\end{split}
\label{rs15}
\end{equation}

In this work, we are going to use the KEH method shown in \cite{komatsu1989rapidly} in order
to solve these equations. This method consists mainly in using the appropriated Green functions to transform the equations $(\ref{rs13})$, $(\ref{rs14})$ and $(\ref{rs15})$ into integral equations, in the following way:
\begin{equation}
\rho =-e^{-\gamma /2}\sum_{n=0}^{\infty }P_{2n}\left( \mu \right)
\int_{0}^{\infty }r^{\prime 2}f_{2n}^{2}\left( r,r^{\prime
}\right) \int_{0}^{1}P_{2n}\left( \mu ^{\prime }\right)
S_{\rho }\left( r^{\prime },\mu ^{\prime }\right)d\mu^{\prime } dr^{\prime},  \label{rs16}
\end{equation}

\begin{equation}
\gamma  =-\frac{2e^{-\gamma /2}}{\pi r\sin \theta }\sum_{n=1}^{\infty }%
\frac{\sin{\left( 2n-1\right) \theta} }{2n-1}\int_{0}^{\infty }r^{\prime 2}f_{2n-1}^{1}\left( r,r^{\prime }\right) \int_{0}^{1}\sin{\left( 2n-1\right) \theta^{\prime}}S_{\gamma }\left( r^{\prime },\mu ^{\prime }\right)d\mu^{\prime } dr^{\prime},
\label{rs17}
\end{equation}

\begin{equation}
\begin{split}
\omega & =-\frac{e^{\left( 2\rho -\gamma \right) /2}}{r\sin \theta }%
\sum_{n=1}^{\infty }\frac{P_{2n-1}^{1}\left( \mu \right) }{2n\left(
2n-1\right) }\int_{0}^{\infty }r^{\prime 3}f_{2n-1}^{2}\left(
r,r^{\prime }\right) \int_{0}^{1}P_{2n-1}^{1}\left( \mu ^{\prime }\right)
S_{\omega }\left( r^{\prime },\mu ^{\prime }\right) d\mu^{\prime } dr^{\prime},
\end{split}
\label{rs18}
\end{equation}
where $P_{n}$ are the Legendre polynomials, $P^{m}_{n}$ are the Associated Legendre functions and $f^{1}_{n}$ and $f^{2}_{n}$ are given by
\begin{equation}
f_{n}^{1}\left( r,r^{\prime }\right) =\left\{
\begin{tabular}{l}
${\left( r^{\prime }/r\right) ^{n}\text{, if }r^{\prime }\leq r,}$ \\
${\left( r/r^{\prime }\right) ^{n}\text{, if }r^{\prime }>r,}$%
\end{tabular}
\right.  \label{rs19}
\end{equation}

\begin{equation}
f_{n}^{2}\left( r,r^{\prime }\right) =\left\{
\begin{tabular}{l}
${r^{\prime n}/r^{n+1}\text{, if }r^{\prime }\leq r,}$ \\
${r^{n}/r^{\prime n+1}\text{, if }r^{\prime }>r}$.
\end{tabular}
\right.  \label{rs20}
\end{equation}

An useful feature of the KEH method is that the asymptotic flatness condition for the potentials $\rho$, $\gamma$ and $\omega$ will be readily satisfied provided that the functions $S_{\rho}$, $S_{\gamma}$ and $S_{\omega}$ be well behaved.
We also will have an equation for the $\alpha$ potential, that is given by
\begin{equation}
\begin{split}
\alpha _{,\mu }=& -\nu _{,\mu }-\left\{ \left( 1-\mu
^{2}\right) \left( 1+rB^{-1}B_{,r}\right) ^{2}+\left[ \mu -\left(
1-\mu ^{2}\right) B^{-1}B_{,\mu }\right] ^{2}\right\} ^{-1} \\
& \left[ \frac{1}{2}B^{-1}\left\{ r^{2}B_{,rr}-\left[ \left(
1-\mu ^{2}\right) B_{,\mu }\right] _{,\mu }-2\mu
B_{,\mu }\right\} \left[ -\mu +\left( 1-\mu ^{2}\right)
B^{-1}B_{,\mu }\right] \right.  \\
& +rB^{-1}B_{,r}\left[ \frac{1}{2}\mu +\mu rB^{-1}B_{,r}+%
\frac{1}{2}\left( 1-\mu ^{2}\right) B^{-1}B_{,\mu }\right]  \\
& +\frac{3}{2}B^{-1}B_{,\mu }\left[ -\mu ^{2}+\mu \left( 1-\mu
^{2}\right) B^{-1}B_{,\mu }\right] -\left( 1-\mu ^{2}\right)
rB^{-1}B_{,\mu r} \\
&  (1+rB^{-1}B_{,r})-\mu r^{2}\left( \nu _{,r}\right)
^{2}-2\left( 1-\mu ^{2}\right) r\nu _{,\mu }\nu _{,r}+\mu
\left( 1-\mu ^{2}\right) \left( \nu _{,\mu }\right) ^{2} \\
& -2\left( 1-\mu ^{2}\right) r^{2}B^{-1}B_{,r}\nu _{,\mu
}\nu _{,r}+\left( 1-\mu ^{2}\right) B^{-1}B_{,\mu }\left[
r^{2}\left( \nu _{,r}\right) ^{2}\right. \\
&  \left.-\left( 1-\mu ^{2}\right) \left(
\nu _{,\mu }\right) ^{2}\right] +\left( 1-\mu ^{2}\right) B^{2}e^{-4\nu }\left\{ \frac{1}{4}\mu
r^{4}\left( \omega _{,r}\right) ^{2}+\frac{1}{2}\left( 1-\mu
^{2}\right) r^{3}\omega _{,\mu }\omega _{,r}\right.  \\
& -\frac{1}{4}\mu \left( 1-\mu ^{2}\right) r^{2}\left( \omega _{,\mu }\right) ^{2}+\frac{1}{2}\left( 1-\mu ^{2}\right)
r^{4}B^{-1}B_{,r}\omega _{,\mu }\omega _{,r} \\
& \left. \left. -\frac{1}{4}\left( 1-\mu ^{2}\right) r^{2}B^{-1}B_{,\mu }\left[ r^{2}\left( \omega _{,r}\right) ^{2}-\left( 1-\mu
^{2}\right) \left( \omega_{,\mu}\right) ^{2}\right] \right\} %
\right],
\end{split}
\label{rs21}
\end{equation}
where 
\begin{equation}
    B=e^{\gamma }=e^{\beta +\nu }. \label{rs22}
\end{equation}

To integrate equation $(\ref{rs21})$ for the function $\alpha(r,\mu)$ we use the boundary condition
\begin{equation}
    \alpha(r,1)=\beta(r,1), \label{rs23}
\end{equation}
that can be found by using the requirement of flatness on the axis of rotation. The function $\alpha(r,\mu)$ will satisfy the asymptotic flatness condition because the other potentials already satisfy this condition \cite{komatsu1989rapidly}.
In addition to the field equations, we also need the equation of hydrostatic equilibrium in order to have a complete set of expressions that describes our system. This equation is found by imposing the conservation of the energy-momentum tensor
with
\begin{equation}
    T_{\;\;\mu;\nu}^{\nu}=0. \label{rs24}
\end{equation}

Computing the above equation with the energy-momentum tensor of equation $(\ref{rs2})$ we encounter 
\begin{equation}
\frac{\nabla p}{\left( \varepsilon +p\right) }=\nabla \ln U^{t}-U^{t}U_{\phi
}\nabla \Omega.  \label{rs25}
\end{equation}

To integrate this equation we need a law of rotation that tells us how the star behaves. With this purpose we are going to work with an uniform rotation that means a constant value to the angular velocity
$\Omega$, so that, $\nabla \Omega=0$. In order to describe the stellar matter
we also need an equation of state (EoS), and in this paper we will
employ a polytropic one of the same type of the one 
used in \cite{tooper1964general,butterworth1976structure,komatsu1989rapidly}

\begin{equation}
p=K\varepsilon ^{1+1/N},  \label{rs26}
\end{equation}
where $K$ is a constant and $N$ is the polytropic index. With these considerations we can integrate equation $(\ref{rs25})$ in order to find
\begin{equation}
\left( K\varepsilon ^{1/N}+1\right) ^{\left( N+1\right) }e^{\nu }\sqrt{%
1-v^{2}}=C,  \label{rs27}
\end{equation}
where $C$ is a constant.

So that we can solve numerically the set of equations obtained above we have to fix two parameters \cite{komatsu1989rapidly}. One of these parameters may be the central energy density $\epsilon_{c}$, or alternatively, the ratio of the maximum pressure to the maximum energy density $\kappa=p_{max}/\epsilon_{max}$. The other parameter we may fix is the axis ratio $q=R_{p}/R_{e}$, where $R_{p}$ is the polar radius and $R_{e}$ is the equatorial radius, or alternatively, we may fix the angular velocity $\Omega$.

In the next section we present the modification in GR that was proposed by Peter Rastall in 1972, that
will be considered in the calculations of Sec. IV.

\section{Rastall gravity}

In his proposal of generalization of Einstein's theory of gravity \cite{rastall1972generalization}, Rastall abandons the assumption of conservation of the energy-momentum tensor in curved space-times, equation $(\ref{rs2})$, and proposes that this non-conservation is related to the space-time curvature, so that the following equation is suggested:
\begin{equation}
    T_{\;\;\mu;\nu}^{\nu}=\lambda^{\prime}R_{,\mu},
    \label{rg1}
\end{equation}
where, $\lambda^{\prime}$ is a constant, and $R$ is the Ricci scalar. For convenience, we chose to rewrite the constant  $\lambda^{\prime}$ as follows: $\lambda^{\prime}=\lambda/(k\left(4\lambda-1\right))$, where $\lambda$ and $k$ are also constants. From the above equation we can see that $\lambda$ cannot assume the value $\lambda=1/4$, because at this point we have a divergence. In \cite{visser2018rastall}, it was alleged that Rastall gravity is completely equivalent to Einstein gravity, however, in his comparison of the two theories, Darabi \textit{et al.} \cite{darabi2018einstein} have shown that such allegation is not correct, in fact, other works support that RG and Rastall gravity are not equivalent, such as, \cite{hansraj2019impact,lobo2020thin}.  It is straightforward to show that the above expression is consistent with the field equations
\begin{equation}
  R_{\mu\nu}-\frac{1}{2}g_{\mu\nu}R=k\left( T_{\mu\nu}-\frac{\lambda}{k\left(4\lambda-1\right)}g_{\mu\nu}R\right), \label{rg2}
\end{equation}
and
if we take the trace of this expression we find
\begin{equation}
   R=k\left(4\lambda-1\right)T.
   \label{rg3}
\end{equation}

Therefore, we can rewrite equation $(\ref{rg1})$ in the following way

\begin{equation}
    T_{\;\;\mu;\nu}^{\nu}=\lambda T_{,\mu},
    \label{rg4}
\end{equation}
and after substituting the relation shown in $(\ref{rg3})$ in eq. $(\ref{rg2})$, we get
\begin{equation}
     R_{\mu\nu}-\frac{1}{2}g_{\mu\nu}R=k\tau_{\mu \nu},
     \label{rg5}
 \end{equation}
where
 \begin{equation}
     \tau_{\mu\nu}=T_{\mu\nu}-\lambda g_{\mu\nu}T.
      \label{rg6}
 \end{equation}
 
 We can notice that equation $(\ref{rg5})$ is equivalent to Einstein field equations but with an effective energy-momentum tensor. It is also interesting to notice that the effective energy-momentum tensor is conserved, that is
 \begin{equation}
     \tau_{\;\;\mu;\nu}^{\nu}=0.
     \label{rg7}
 \end{equation}

 In order to determine the value of the constant $k$,
 in the next section, we are going to take the Newtonian limit of Rastall field equations.

\subsection{Newtonian Limit}

In this section, we take the Newtonian limit of equation $(\ref{rg2})$ and end up finding the value of the constant $k$. In this limit, we want to recover the Poisson equation, that is given by
\begin{equation}
    \nabla^{2}\phi=4\pi \rho,
    \label{nl1}
\end{equation}
where $\phi$ is the gravitational potential and $\rho$ is the mass density. In the weak field and low velocity regime, the metric tensor $g_{\mu\nu}$ can be replaced by the Minkowski tensor $\eta_{\mu\nu}$. Also in this regime, the energy density $T_{tt}$ is equal to the mass density $\rho$, so that the $(tt)$ component of equation $(\ref{rg2})$ will be written in the following way
\begin{equation}
    R_{tt}+\frac{2\lambda-1 }{2\left(4\lambda-1\right)}R=-k\rho. \label{nl2}
\end{equation}

Using the approximation $R\approx R_{ii}-R_{tt}$, we can find the relation
\begin{equation}
    R=2\frac{4\lambda-1}{2\lambda+1}R_{tt}, \label{nl3}
\end{equation}
and
replacing it in equation $(\ref{nl2})$, we find
\begin{equation}
    R_{tt}=-\frac{1}{2}\left(2\lambda+1\right)k \rho. \label{nl4}
\end{equation}

Finally, knowing that  $R_{tt}\approx-\nabla^{2}\phi$ \cite{weinberg1973gravitation}, we can find that the value of $k$ will be given by
\begin{equation}
   k=\frac{8\pi}{2\lambda+1}. \label{nl5}
\end{equation}

We can verify that, as it would be expected, taking the limit of $\lambda\rightarrow 0$, we will recover the value of $k$ in GR, that is, $k=8\pi$. Equation $(\ref{nl5})$ also show us that besides $\lambda=1/4$, we also cannot have $\lambda=-1/2$ because at this point there is also a divergence.

In the next section, we are going to show how the equations for a rapidly rotating star presented in Section 2 are modified by Rastall's theory.

\section{Rotating stars in Rastall gravity}

In this section we show how the equations of the KEH method for a rotating star can be written in Rastall's theory. As we can see from equation $(\ref{rg5})$, in order to find the field equations for Rastall's gravity we have to compute the effective energy-momentum tensor. Starting from equation $(\ref{rg6})$ we can show that in this case the effective energy-momentum tensor will be given by
\begin{equation}
     \tau_{\mu\nu}=p_{ef}g_{\mu\nu}+(p_{ef}+\varepsilon_{ef})U_{\mu}U_{\nu},
      \label{rr1}
 \end{equation}
where
 \begin{align}
        p_{ef}=&p\left(1-3\lambda\right)+\lambda\varepsilon, \\
     \varepsilon_{ef}=&\varepsilon\left(1-\lambda\right)+3\lambda p,
     \label{rr2}
 \end{align}
 and the metric tensor $g_{\mu\nu}$ and four-velocity $U_{\mu}$ are the same ones used in GR, given respectively by equations $(\ref{rs1})$ and $(\ref{rs3})$. Now if we calculate the equations $(\ref{rg5})$ we are going to conclude that the equation $(\ref{rs21})$, for the potential $\alpha$, will still be the same because this equation does not depend on the energy-momentum tensor. On the other hand, the equations $(\ref{rs13})$, $(\ref{rs14})$ and $(\ref{rs15})$ for the potentials $\rho$,
$\gamma$ and $\omega$ will be almost the same, the only difference being that now the functions $S_{\rho}$, $S_{\gamma}$ and $S_{\omega}$ are going to be modified in the following way:
\begin{equation}
\begin{split}
S_{\rho }(r,\mu )=& \left[ e^{\gamma /2} k e^{2\alpha }\left( \varepsilon
+p\right) \frac{1+v^{2}}{1-v^{2}}\right.  \\
& +r^{2}(1-\mu ^{2})e^{-2\rho }\left[ \omega _{,r}^{2}+\frac{1}{%
r^{2}}\left( 1-\mu ^{2}\right) \omega _{,\mu }^{2}\right] +\frac{1}{%
r}\gamma _{,r}-\frac{1}{r^{2}}\mu \gamma _{,\mu } \\
& \left. +\frac{\rho }{2}\left\{ 2 k e^{2\alpha }p_{ef}-\gamma _{,r}\left( \frac{1}{2}\gamma _{,r}+\frac{1}{r}\right) -\frac{1}{r^{2}%
}\gamma _{,\mu }\left[ \frac{1}{2}\gamma _{,\mu }\left(
1-\mu ^{2}\right) -\mu \right] \right\} \right],
\end{split}
\label{rr3}
\end{equation}

\begin{equation}
S_{\gamma }\left( r,\mu \right) =e^{\gamma /2}\left\{ 2 k e^{2\alpha }p_{ef}+%
\frac{\gamma }{2}\left[ 2k e^{2\alpha }p_{ef}-\frac{1}{2}\gamma _{,r}^{2}-\frac{1}{2r^{2}}\left( 1-\mu ^{2}\right) \frac{1}{2}\gamma
_{,\mu }^{2}\right] \right\},  \label{rr4}
\end{equation}
and
\begin{equation}
\begin{split}
S_{\omega }\left( r,\mu \right) =& e^{\left( \gamma -2\rho \right) /2}\left[
-2 k e^{2\alpha }\frac{\left( \Omega -\omega \right) \left( \varepsilon
+p\right) }{1-v^{2}}\right.  \\
& +\omega \left\{ - k e^{2\alpha }\left[\frac{ \left( 1+v^{2}\right)
\left(\varepsilon +p\right)}{1-v^{2}}-p_{ef}\right] -\frac{1}{r}\left( 2\rho _{,r}+\frac{1}{2}\gamma _{,r}\right) \right.  \\
& +\frac{1}{r^{2}}\mu \left( 2\rho _{,\mu }+\frac{1}{2}\gamma
_{,\mu }\right) +\frac{1}{4}\left( 4\rho _{,r}^{2}-\gamma
_{,r}^{2}\right) +\frac{1}{4r^{2}}\left( 1-\mu ^{2}\right) \left(
4\rho _{,\mu }^{2}-\gamma _{,\mu }^{2}\right)  \\
& \left. \left. -r^{2}\left( 1-\mu ^{2}\right) e^{-2\rho }\left[ \omega
_{,r}^{2}+\frac{1}{r^{2}}\left( 1-\mu ^{2}\right) \omega _{,\mu }^{2}\right] \right\} \right].
\end{split}
\label{rr5}
\end{equation}

Here it is interesting to notice that $p_{ef}+\varepsilon_{ef}=p+\varepsilon$, so that we can change from the sum of the effective functions to the sum of the usual ones when it is more convenient. Now we compute the law of conservation from equation $(\ref{rg4})$, so that we obtain the relation
\begin{equation}
\frac{\nabla p_{ef}}{\left( \varepsilon +p\right) }=\nabla \ln u^{t}-u^{t}u_{\phi
}\nabla \Omega.  \label{rr6}
\end{equation}

If we once again assume uniform rotation and a polytropic EoS, the integration of the above equation will result in
\begin{equation}
\left( N\left(1-4\lambda\right)+\left(1-3\lambda\right)\right) \ln \left( K\varepsilon ^{1/N}+1\right)+\lambda\ln\left(\varepsilon\right) +\nu +\frac{1}{2}%
\ln \left( 1-v^{2}\right) =C, \label{rr7}
\end{equation}
where $K$ and $N$ are the constants from the polytropic EoS, equation $(\ref{rs26})$, and $C$ is a constant of integration.

In this work we will analyse the mass radius relation and for this purpose we are going to calculate the Komar mass \cite{komar1959covariant,wald2010general}, that for stationary, asymptotically flat space-times is the same as the ADM mass \cite{arnowitt1959dynamical,arnowitt1960canonical,friedman2013rotating} and that is also referred as the total mass or gravitational mass in some works \cite{komatsu1989rapidly,cook1992spin}. It is possible to show that the total mass of the star can be given by the following expression in our system:
\begin{equation}
    M=\frac{1}{2\lambda+1}\int\left(-2\tau^{t}_{\;t}+\tau^{\sigma}_{\;\sigma}\right)\sqrt{-g}\;d^{3}x.
\end{equation}

If we compute the above expression for our equations, we obtain
\begin{equation}
    \begin{split}
          M=&\frac{4\pi}{2\lambda+1}\int_0^{\pi/2}\int_0^{R_{e}} e^{2\alpha+\beta}\left[e^{\nu}\left(\frac{\left(\epsilon+p\right)\left(1+v^{2}\right)}{1-v^{2}}+2p_{ef}\right)+\right. \\ &\left.2r\sin\theta\;\omega e^{\beta}\frac{\left(\epsilon+p\right)v}{1-v^{2}}\right]r^{2}\sin\theta\; dr d\theta,  
    \end{split}
\end{equation}
where $R_{e}$ is the equatorial radius. It is important to notice that the radial coordinate of our metric, given by equation $(\ref{rs1})$, is not the same as the radial coordinate of the Schwarzschild metric \cite{butterworth1976structure}. The equatorial radius in the Schwarzschild metric $R^{(Sch)}_{e}$  is related to the equatorial radius $R_{e}$ of our metric by the expression 
\begin{equation}
    R^{(Sch)}_{e}=R_{e}\left(1+\frac{M}{2R_{e}}\right)^{2}.
\end{equation}

This relation is important so that we can correctly compare our results with those obtained for a static star. We will also analyse the moment of inertia $I$, that can be given by the ratio between the angular momentum $J$ and the angular velocity $\Omega$,
\begin{equation}
    I=\frac{J}{\Omega},
\end{equation}
where the angular momentum $J$ in our case is given by 
\begin{equation}
    J=\frac{1}{2\lambda+1}\int \tau^{t}_{\;\phi}\sqrt{-g}\;d^{3}x.
\end{equation}

If we compute this expression with the equations given above we obtain
\begin{equation}
    J=\frac{4\pi}{2\lambda+1}\int_0^{\pi/2}\int_0^{R_{e}} \frac{\left(\epsilon+p\right)v}{1-v^{2}} e^{2\alpha+2\beta} r^{3}\sin^{2}\theta\; dr d\theta.
\end{equation}

 We have performed our calculations for two types of relevant sequences of solutions, one for the set of static stars and the other one for the set of stars rotating at the mass shedding limit, also called Kepler limit. This limit is reached when the angular velocity of an element of fluid at the star boundary at the equator $\Omega(R_{e},\pi/2)$ is the same as the angular velocity of a free particle in circular orbit $\Omega_{K}$ in this same point. The Kepler limit $\Omega_{K}$ is given by the relation \cite{friedman2013rotating}
 \begin{equation}
     \Omega_{K}=\left.\left(\omega+\frac{r\partial_{r}\omega}{2(1+r\partial_{r}\beta)}+\sqrt{\left(\frac{r\partial_{r}\omega}{2(1+r\partial_{r}\beta)}\right)^{2}+\frac{e^{2(\nu-\beta)}\partial_{r}\nu}{r(1+r\partial_{r}\beta)}}\;\right)\right|_{r=R_{e},\theta=\frac{\pi}{2}}.
 \end{equation}

\section{Numerical method}

We have constructed our numerical code following the method shown in \cite{komatsu1989rapidly}, with the implementation of the equations modified by Rastall's gravity. We also implemented a modification suggested in \cite{stergioulas1994comparing,stergioulas1996structure} that improves the solution of the potential $\alpha$. The numerical procedure starts by providing an initial approximation for the potentials and the energy density, in order to obtain this approximation we have followed the approach presented in \cite{butterworth1976structure}. A more detailed description of the KEH method can be found in \cite{komatsu1989rapidly,cook1992spin,friedman2013rotating}. For obtaining the solutions for the static sequences we have implemented a modification of the same code used in \cite{mota2019combined}, that is based on Runge-Kutta method.

We have checked our results for the GR limit by setting $\lambda=0$ and comparing with the correspondent results, and we also checked the non-rotating limit by making $\Omega \rightarrow 0$ and comparing our results with the ones for static stars. In both cases we have encountered great agreement between the solutions. 

\section{Results}

In this section we are going to analyse the numerical results of our equations for rotating stars in Rastall gravity. We will work with two polytropic equations of state, given by equation $(\ref{rs26})$. In the first one, that we will refer to by $EoS_1$, we have chosen $N_{1}=1.5$ as in \cite{stoeckly1965polytropic,ray2003electrically} and $K_{1}=125269.51$, and in the second one, $EoS_2$, we choose $N_{2}=0.7463$ as in \cite{doneva2013rapidly,kleihaus2016rapidly} and $K_{2}=3.25 \times 10^{11}$, in both cases the constant $K$ is in geometrized units. We have considered four different values of the parameter $\lambda$:  $0$, $5\times10^{-4}$, $1\times10^{-3}$ and $5\times10^{-3}$. When we set $\lambda=0$ we retrieve the equations for GR. We choose not to work with negative values for $\lambda$, as this leads to negative values for $p_{ef}$, which makes the system unstable. In our figures the curves with solid line represent the mass shedding sequences and the dash-dot curves represent the static sequences. 

\begin{figure}
     \centering
     \begin{subfigure}[b]{0.49\textwidth}
         \centering
         \includegraphics[width=\textwidth]{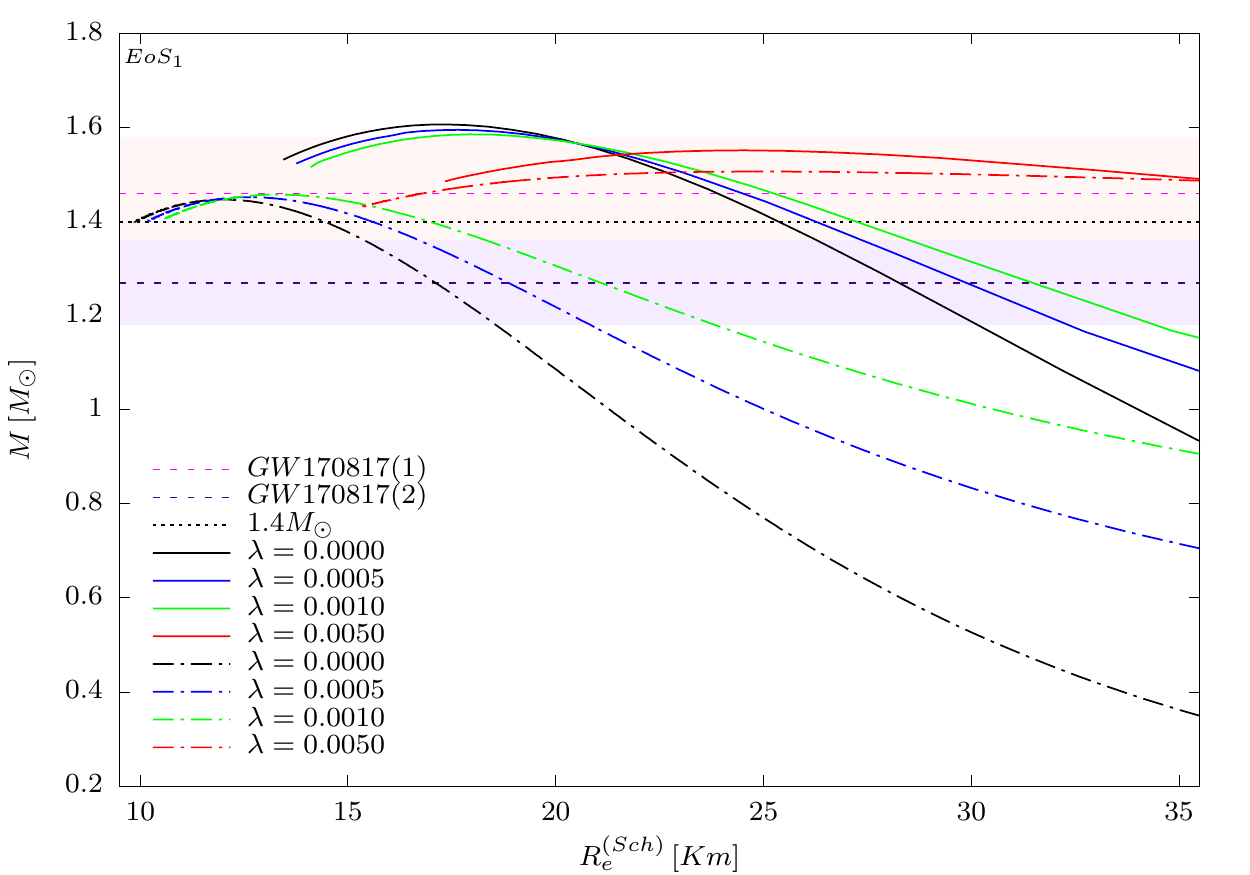}
     \end{subfigure}
     \begin{subfigure}[b]{0.49\textwidth}
         \centering
         \includegraphics[width=\textwidth]{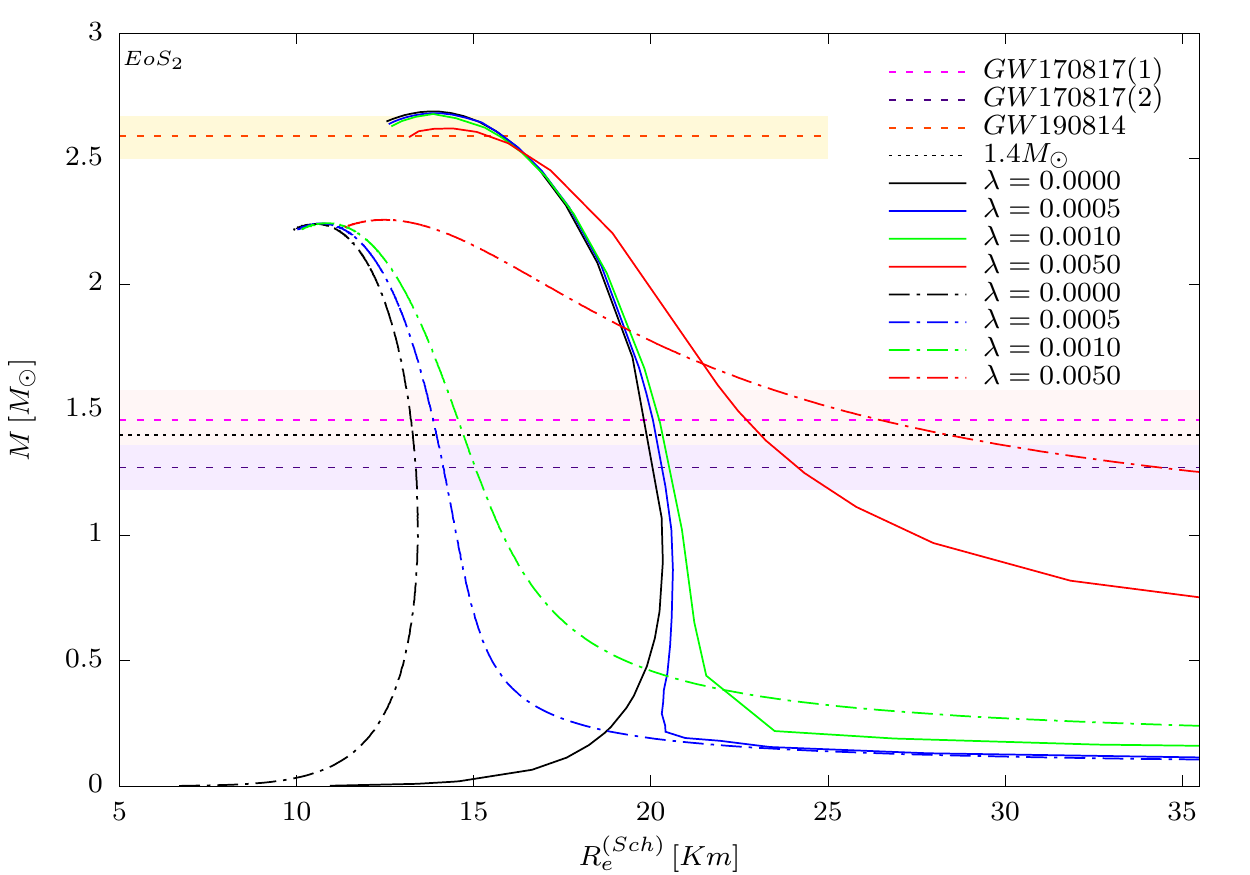}
     \end{subfigure}
        \caption{The mass-radius relation for sequences of non-rotating stars (dash-dot curves) and for stars rotating at the mass-shedding limit (solid line curves). The curves for $\lambda=0$ correspond to the GR case.}
        \label{fig1}
\end{figure}

In Figure $\ref{fig1}$ we show the mass-radius relation, we can see that in all sequences analysed the parameter $\lambda$ affects more intensely the radius of the stars, increasing its value, while the masses are changed more weakly. As already observed, for the non-rotating sequences \cite{fabris2012rastall,mota2019combined}, the maximum mass is lightly increased, but for stars rotating at the mass shedding limit, the effect is the opposite, that is, the maximum mass is lightly diminished. It is possible to notice that the effect of Rastall's modification is more intense on the $EoS_1$, that can be considered a more soft EoS, and also the effect of this parameter is more intense in the static stars. It is also interesting to notice that $EoS_2$ produces a mass-radius curve similar to the curves for quark stars in the GR case, but this same EoS produces curves similar to neutron star mass-radius curves in Rastall's gravity for both the static and rotating stars.

Figure $\ref{fig1}$ also shows the mass of a canonical $1.4 M_{\odot}$ neutron star (black-dotted horizontal line) and the masses of the two neutron stars in the event GW170817 \cite{abbott2017gw170817}, with dashed regions to represent the uncertainty in the mass measures, respectively $1.46_{-0.10}^{+0.12} M_{\odot}$ (pink-dashed horizontal line) and $1.27_{-0.09}^{+0.09} M_{\odot}$ (purple-dashed horizontal line). In the right panel of Figure $\ref{fig1}$ we also show the mass of the compact object detected in the event GW190814 \cite{abbott2020gw190814}, that has a mass of $2.59_{-0.09}^{+0.08} M_{\odot}$ (orange-dotted horizontal line). Recently \cite{landry2020nonparametric}, the data from gravitational waves observations was combined with other sources of information to estimate the radius of $R_{1.4}=12.32_{-1.47}^{+1.09}km$ for a canonical neutron star. Data from LIGO and Virgo \cite{abbott2018gw170817} also have been used to estimate the radii of the stars in the GW170817 event to be $R_1=10.8_{-1.7}^{+2.0}km$ and $R_2=10.7_{-1.5}^{+2.1}km$, respectively. Analysing Figure $\ref{fig1}$, we can observe that the mass-radius curves with $EoS_2$ are in better agreement with this data, since $EoS_1$ provides too big radii for the stars analysed. If we use the mass-radius curve with $EoS_2$ for a static star in Rastall gravity with $\lambda=5 \times 10^{-4}$, to predict the radii of the analysed stars, we will obtain: $R_{1.4} \approx 13.95km$, $R_1 \approx 13.85km$ and $R_2 \approx 14.14km$, respectively. Lastly, we can observe that our mass-radius curves for stars rotating at the Keplerian limit with $EoS_2$, can fit the the compact object within the mass gap from the event GW190814 with a radius of $R \approx 15.79km$ in Rastall gravity with $\lambda=5 \times 10^{-4}$. 

\begin{figure}
     \centering
     \begin{subfigure}[b]{0.49\textwidth}
         \centering
         \includegraphics[width=\textwidth]{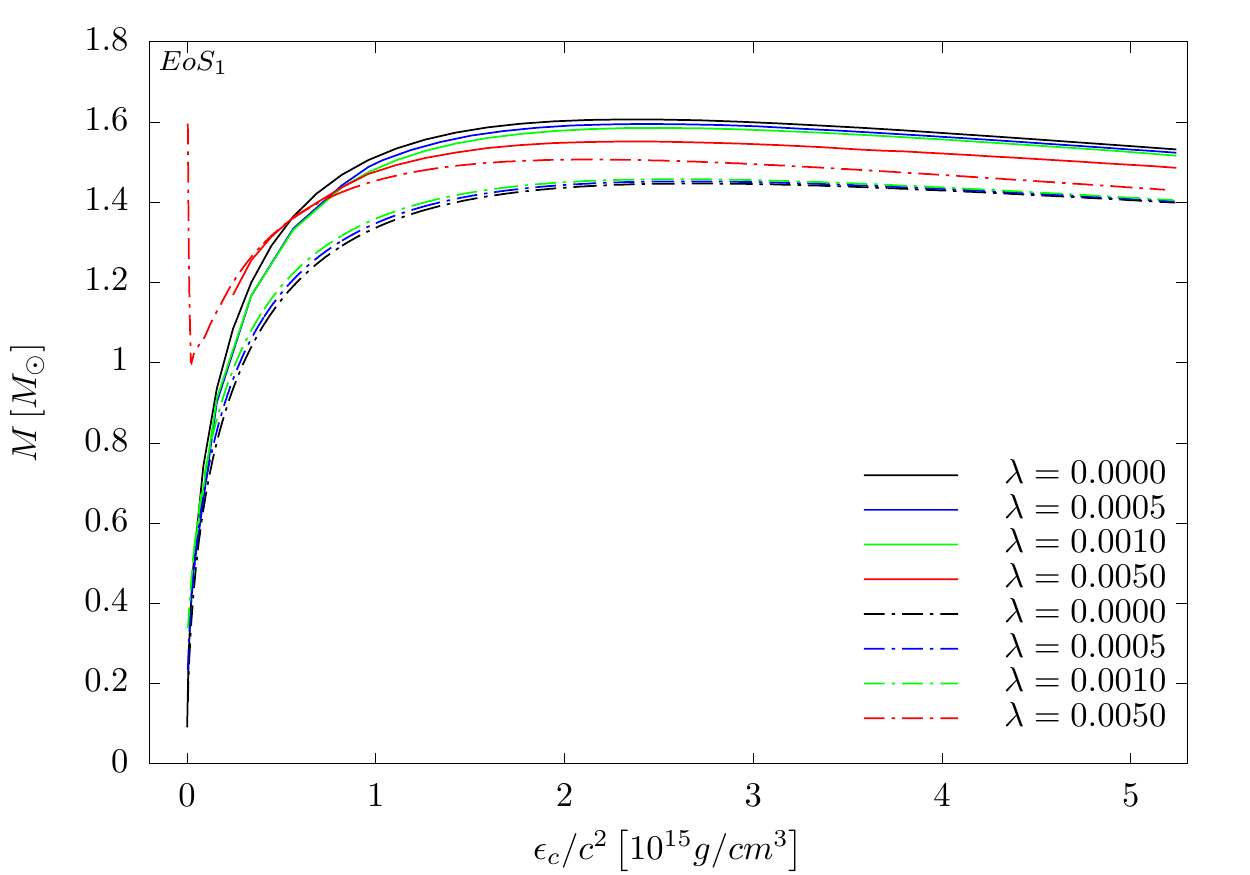}
     \end{subfigure}
     \begin{subfigure}[b]{0.49\textwidth}
         \centering
         \includegraphics[width=\textwidth]{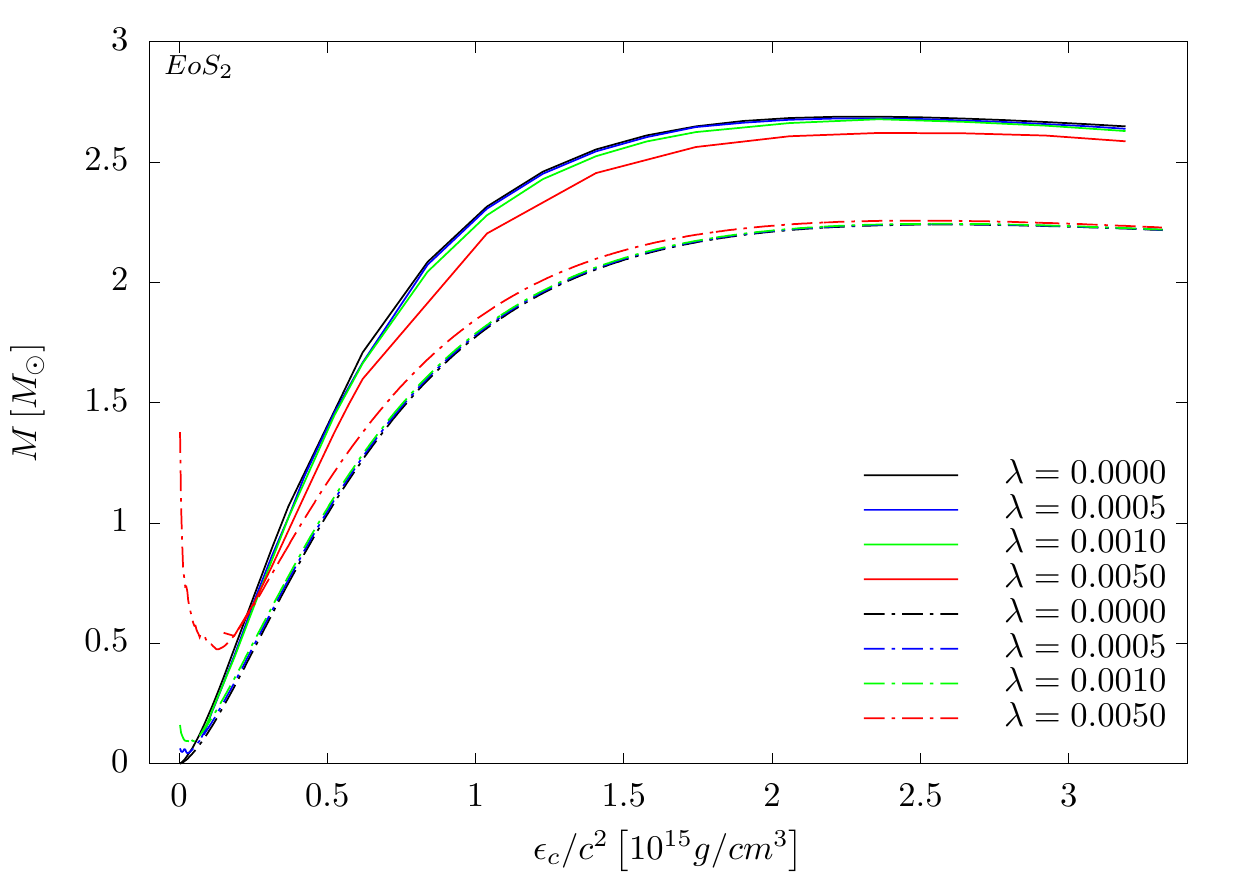}
     \end{subfigure}
        \caption{The mass as a function of the central energy density for sequences of non-rotating stars (dash-dot curves) and for stars rotating at the mass-shedding limit (solid line curves). The curves for $\lambda=0$ correspond to the GR case.}
        \label{fig2}
\end{figure}

\begin{figure}
     \centering
     \begin{subfigure}[b]{0.49\textwidth}
         \centering
         \includegraphics[width=\textwidth]{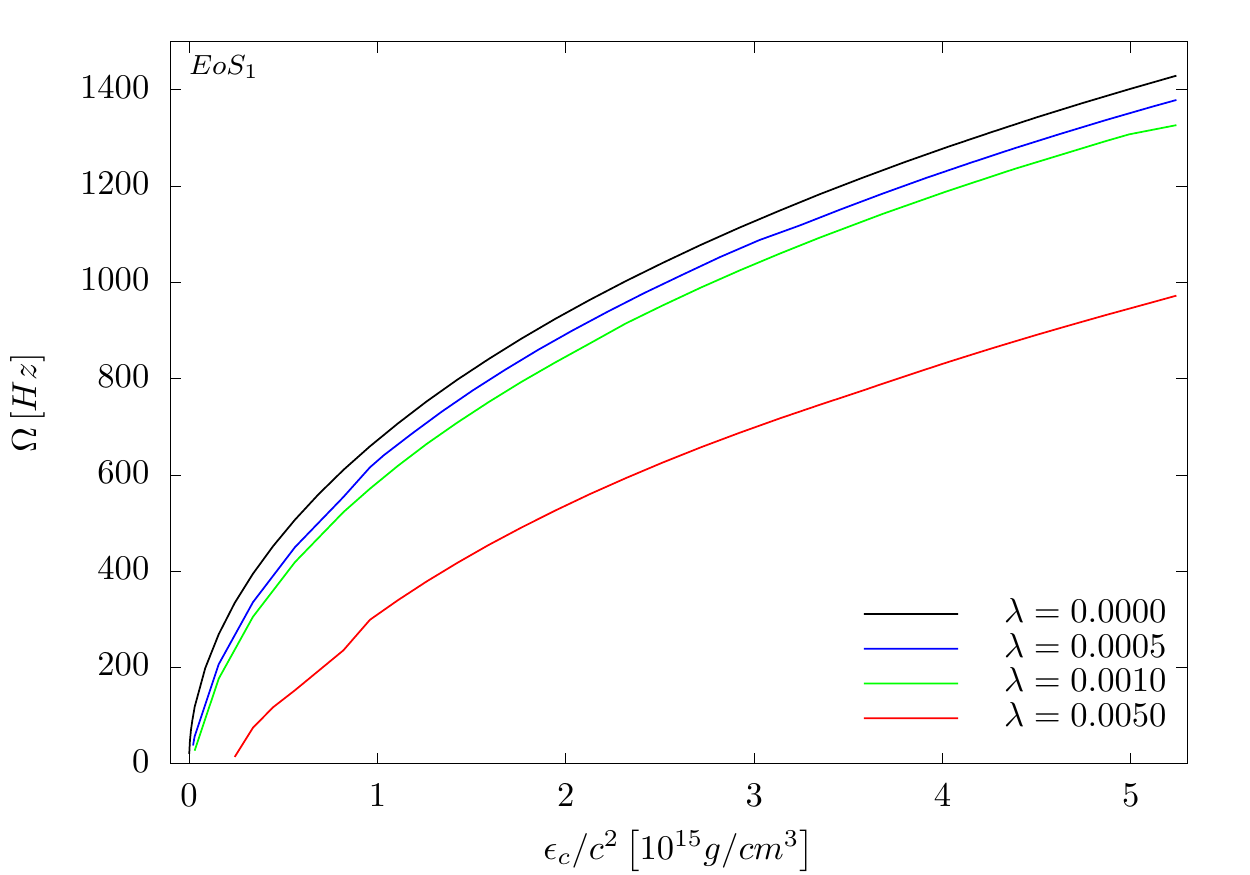}
     \end{subfigure}
     \begin{subfigure}[b]{0.49\textwidth}
         \centering
         \includegraphics[width=\textwidth]{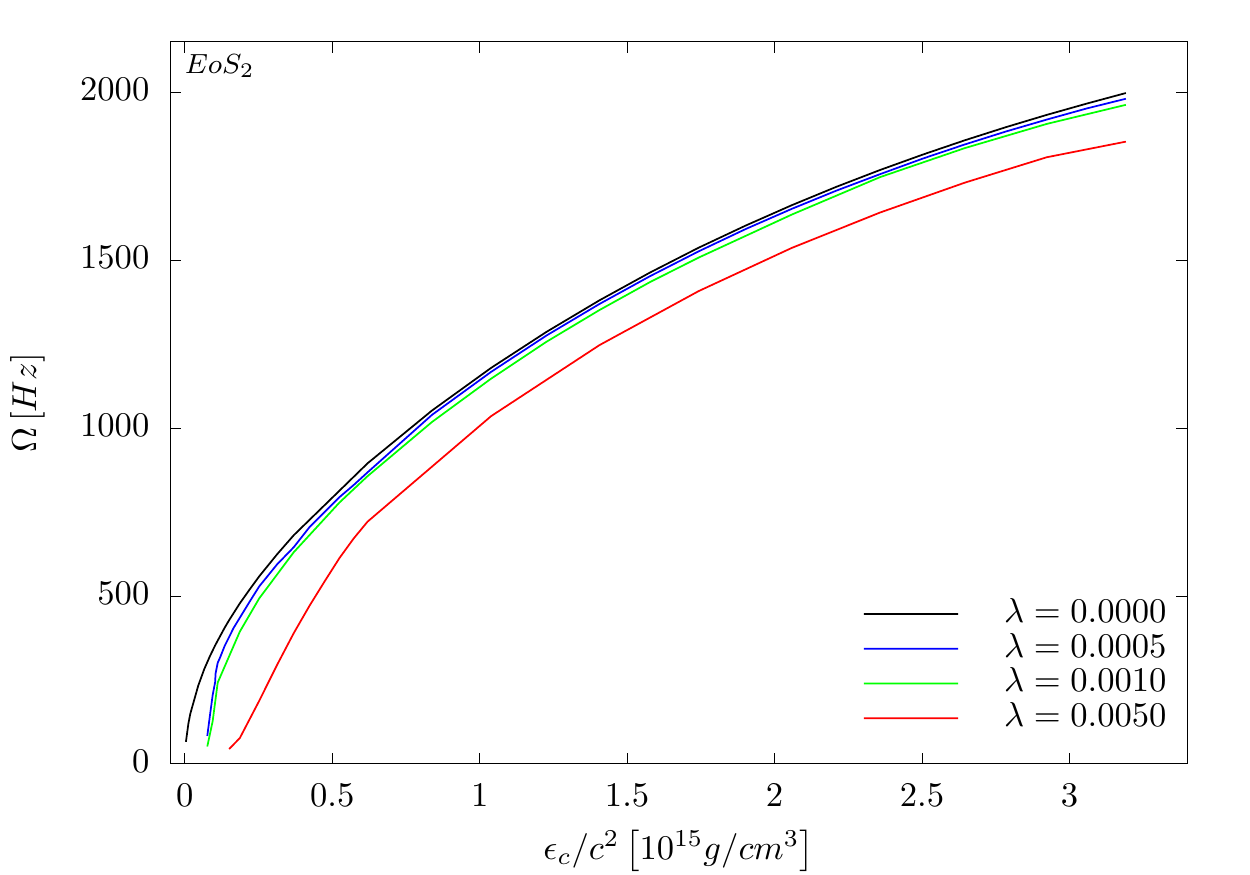}
     \end{subfigure}
        \caption{The angular velocity as a function of the central energy density for sequences of stars rotating at the mass-shedding limit. The curves for $\lambda=0$ correspond to the GR case.}
        \label{fig3}
\end{figure}

In Figure $\ref{fig2}$ we can see the relation between mass and central energy density, this figure also shows that Rastall's gravity has opposite effects in the masses of the static stars and stars at the Kepler limit. For the static sequences an increase in the parameter $\lambda$ produces also an increase in the mass of stars with the same central energy density. But for the mass shedding sequences Rastall's gravity produces stars with smaller masses than GR for the same central energy density.
This can be better understood when we analyse the next figure, Figure \ref{fig3}, where we show the relation between $\Omega_{K}$ and the central energy density. It is possible to observe that the maximum angular velocity with which a star can rotate before starting to lose mass in GR is greater than in Rastall's gravity, for the same central energy density, and the higher the value of $\lambda$ the lower the value of $\Omega_{K}$. 

\begin{figure}
     \centering
     \begin{subfigure}[b]{0.49\textwidth}
         \centering
         \includegraphics[width=\textwidth]{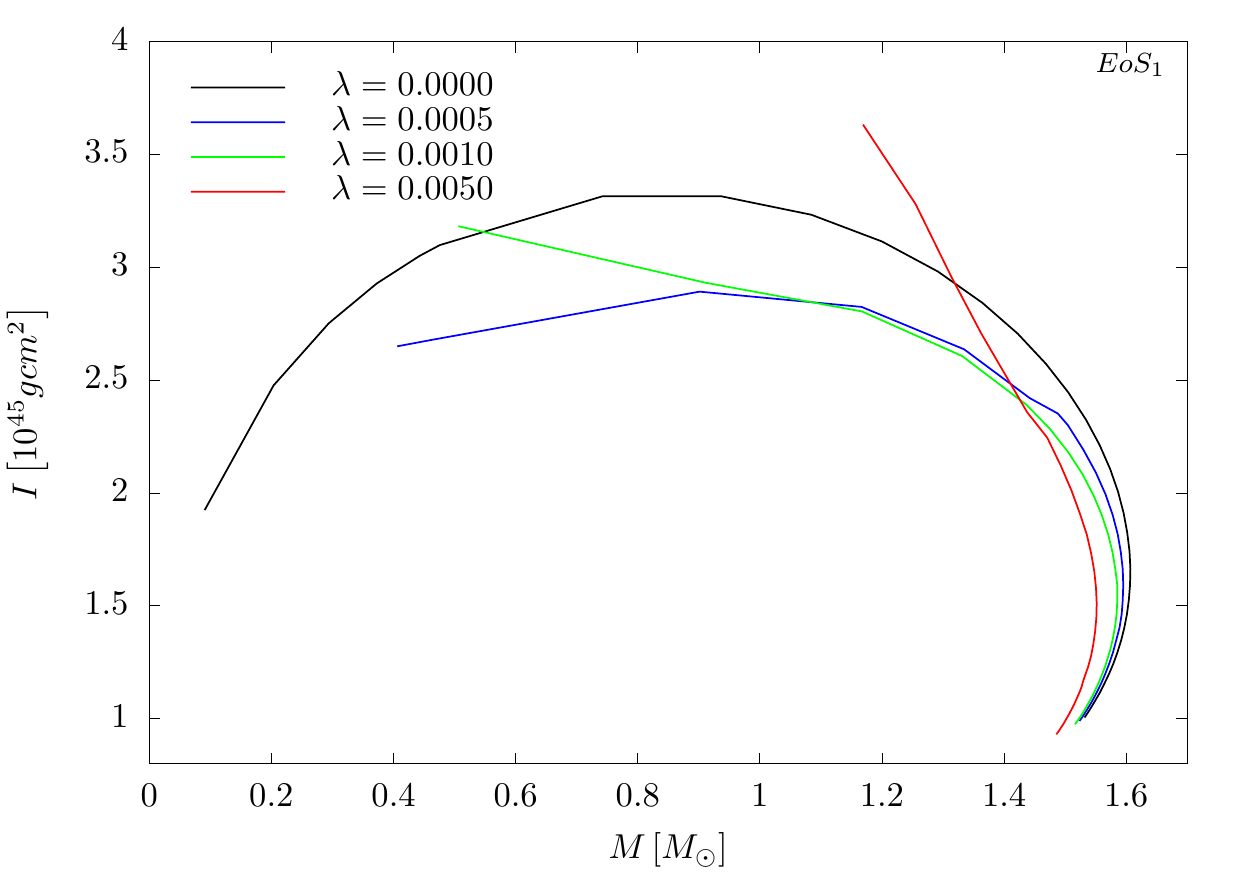}
     \end{subfigure}
     \begin{subfigure}[b]{0.49\textwidth}
         \centering
         \includegraphics[width=\textwidth]{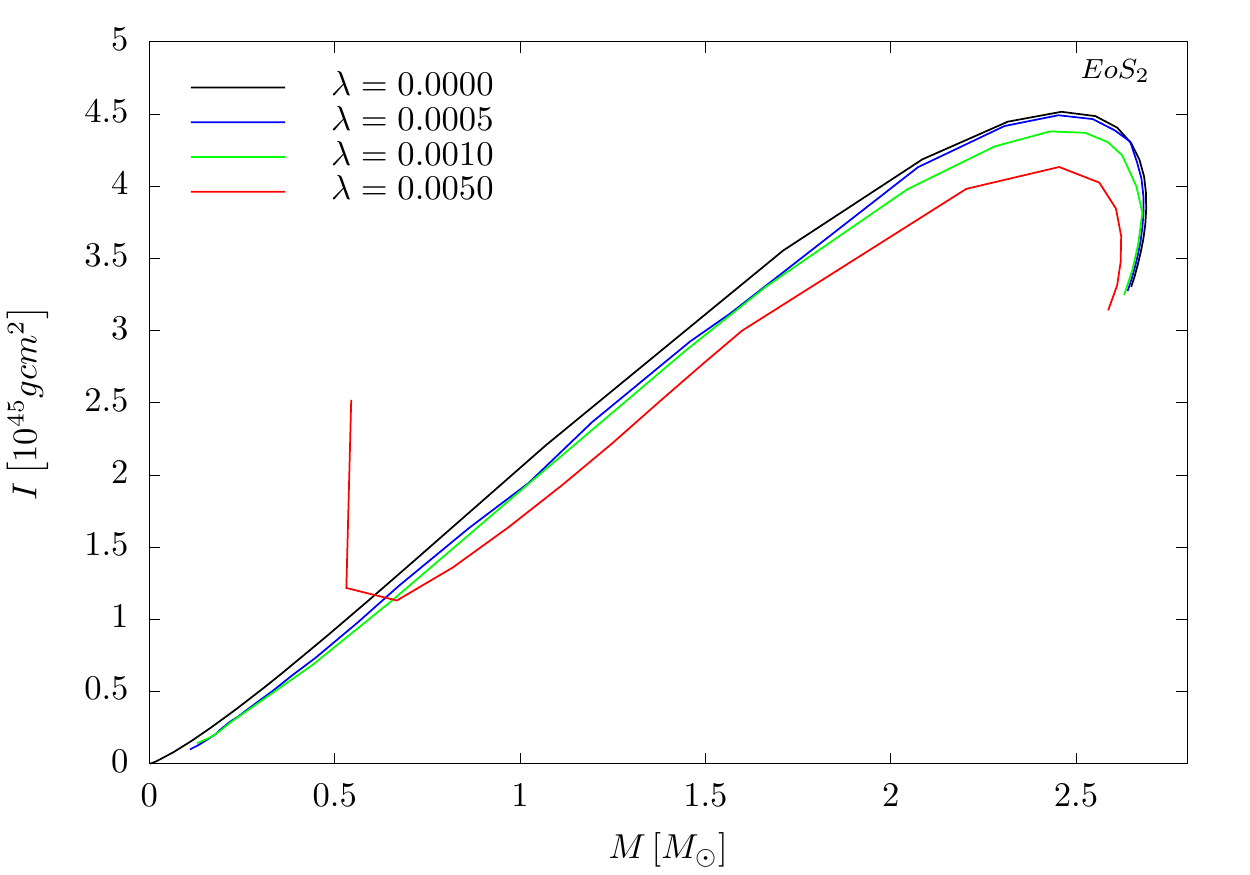}
     \end{subfigure}
        \caption{The Moment of inertia as a function of the mass for sequences of stars rotating at the mass-shedding limit. The curves for $\lambda=0$ correspond to the GR case.}
        \label{fig4}
\end{figure}

The moment of inertia is an important physical quantity to be analyzed, as it can be an effective way of ascertaining the internal structure of the star, that is, its EoS \cite{ozel2016masses}.  Due to the connection between EoS and moment of inertia, it is possible, specially in the case of stiff EoS, to construct approximations for the moment of inertia so that this physical quantity can be estimated also in the study of static models \cite{lattimer2000nuclear,bejger2002moments,tello2019anisotropic,singh2019minimally,singh2020static}. In Figure \ref{fig4} we can see how this quantity is affected by Rastall's gravity. In general, the theory of Rastall diminishes the value of $I$, but for some smaller masses the behaviour of the $I\times M$ curve in Rastall's gravity diverges from the one found in GR. Here we can also note that the modified theory of Rastall has a greater effect on $EoS_1$ than on $EoS_2$.

\begin{figure}
     \centering
     \begin{subfigure}[b]{0.49\textwidth}
         \centering
         \includegraphics[width=\textwidth]{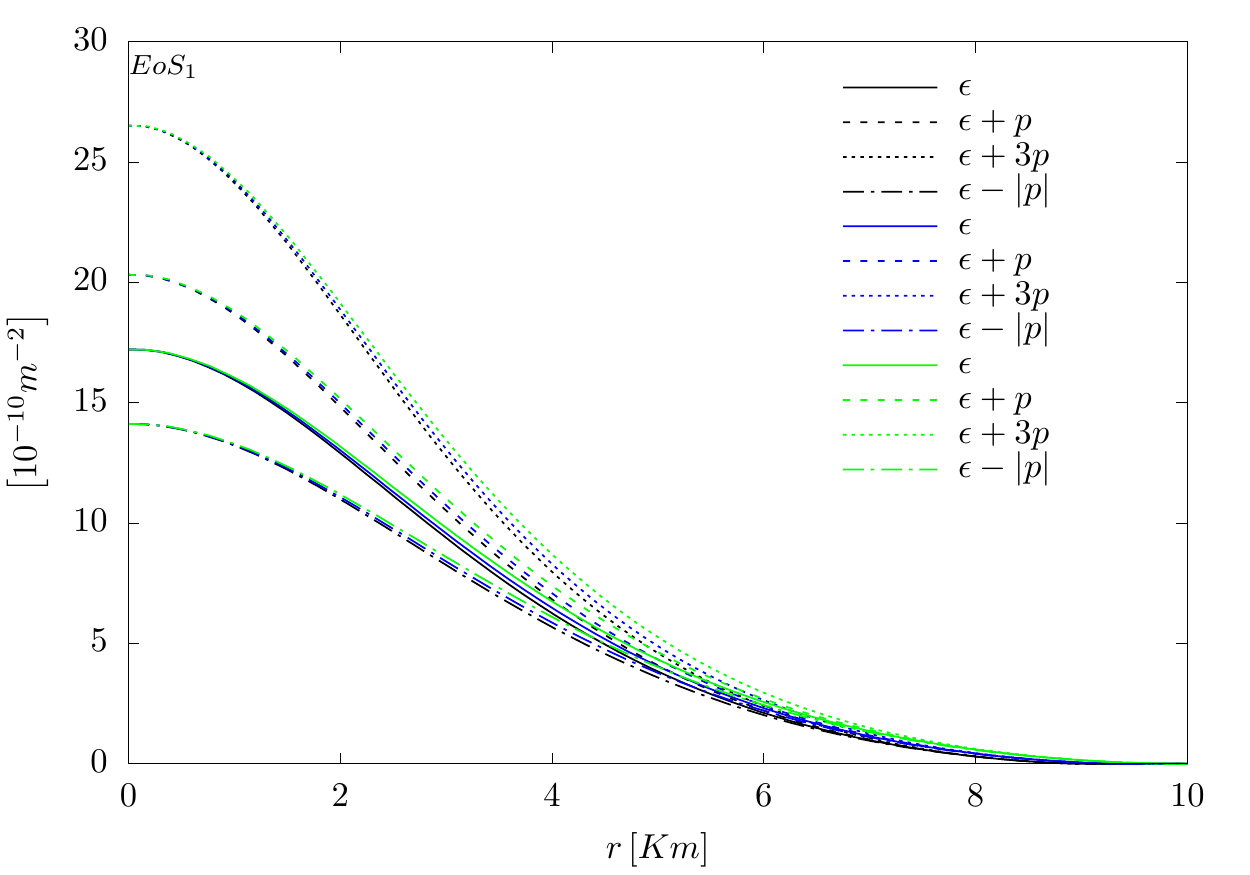}
     \end{subfigure}
     \begin{subfigure}[b]{0.49\textwidth}
         \centering
         \includegraphics[width=\textwidth]{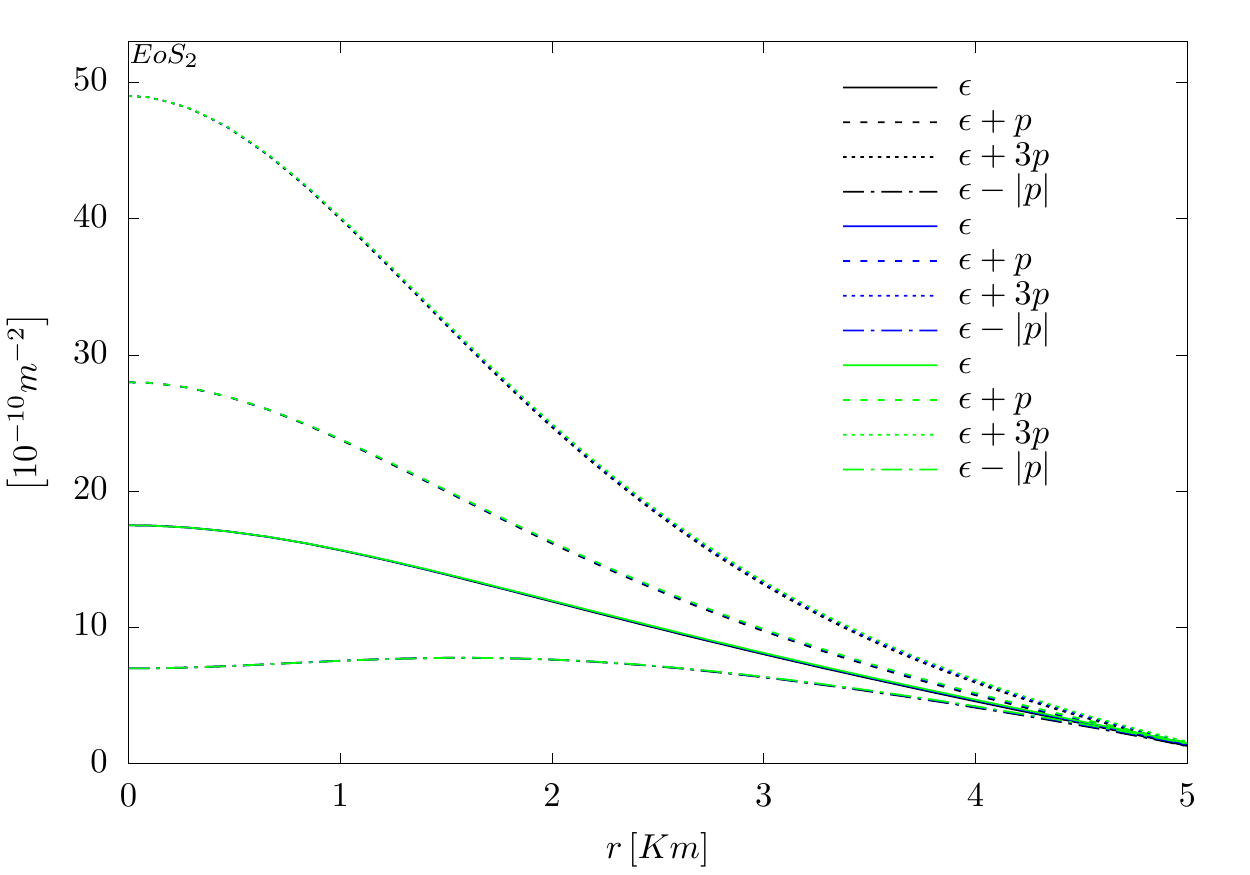}
     \end{subfigure}
        \caption{Variation of energy conditions with radial coordinate $r$ at the equator $(\theta=\pi/2)$, for stars rotating at the mass-shedding limit with a fixed central energy density.}
        \label{fig6}
\end{figure}

Something important to consider to ensure our solutions are physically acceptable is whether the energy conditions are satisfied. For a perfect fluid these conditions are given by \cite{wald2010general}:
   
   \textit{Weak energy condition (WEC):} $\epsilon \geq 0$ and $\epsilon + p \geq 0$,
   
   \textit{Strong energy condition (SEC):} $\epsilon + 3p \geq 0$ and $\epsilon + p \geq 0$,
   
    \textit{Dominant energy condition (DEC):} $\epsilon \geq \lvert p \rvert$.

As we only worked with positive values for $\epsilon$ e $p$, it is expected that the energy conditions will be satisfied. As an example, in Figure \ref{fig6}, we have plotted all the above conditions for stars with central energy density of $\epsilon_{c}/c^{2}=2.3\times 10^{15}g/cm^3$ on the left panel, and central energy density of $\epsilon_{c}/c^{2}=2.4\times 10^{15}g/cm^3$ on the right panel. The energy and pressure are given in geometrized units, the black curves represent the energy conditions in GR, the blue curves are for Rastall gravity with $\lambda=5 \times 10^{-4}$ and the green curves are for Rastall gravity with $\lambda=1 \times 10^{-3}$. We can conclude that all energy conditions are satisfied in Figure \ref{fig6}.

\begin{figure}
     \centering
     \begin{subfigure}[b]{0.49\textwidth}
         \centering
         \includegraphics[width=\textwidth]{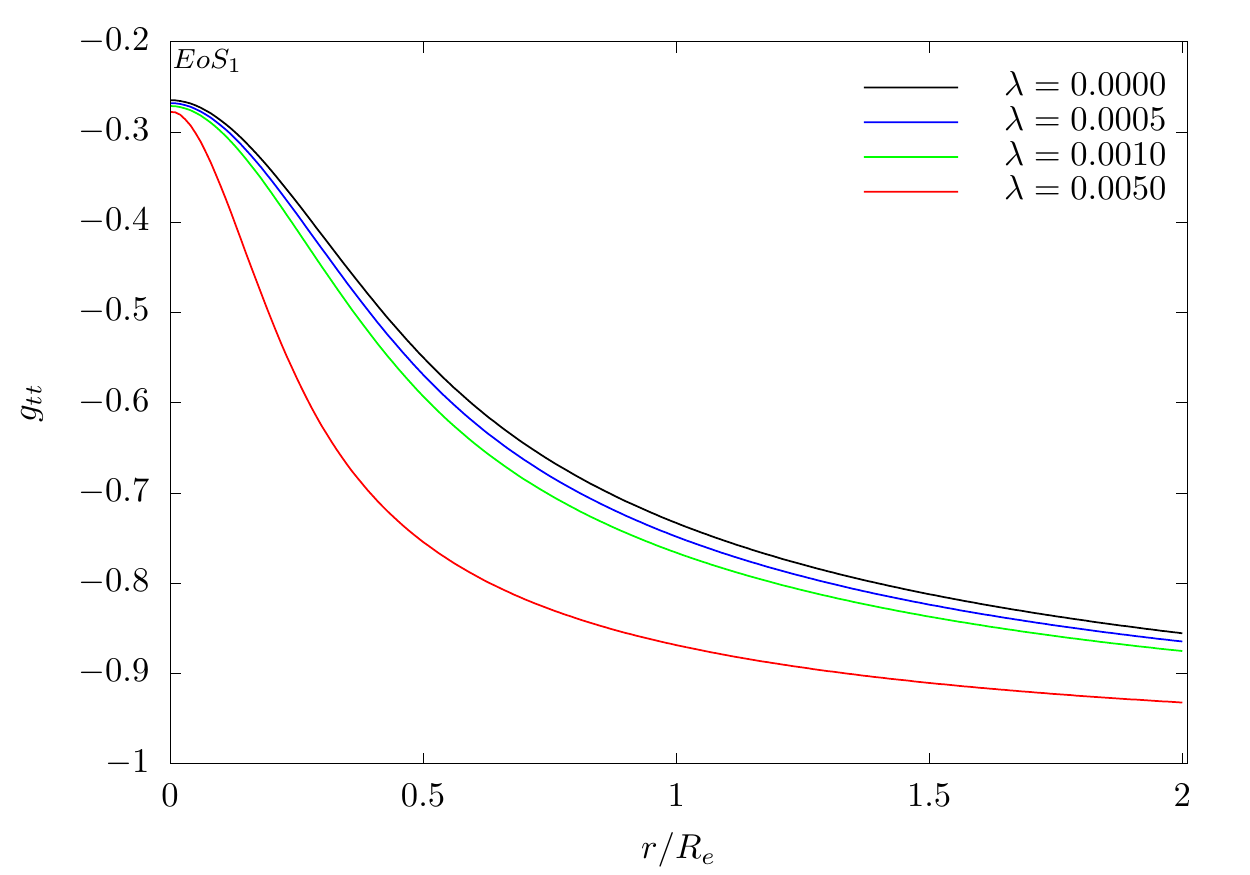}
     \end{subfigure}
     \begin{subfigure}[b]{0.49\textwidth}
         \centering
         \includegraphics[width=\textwidth]{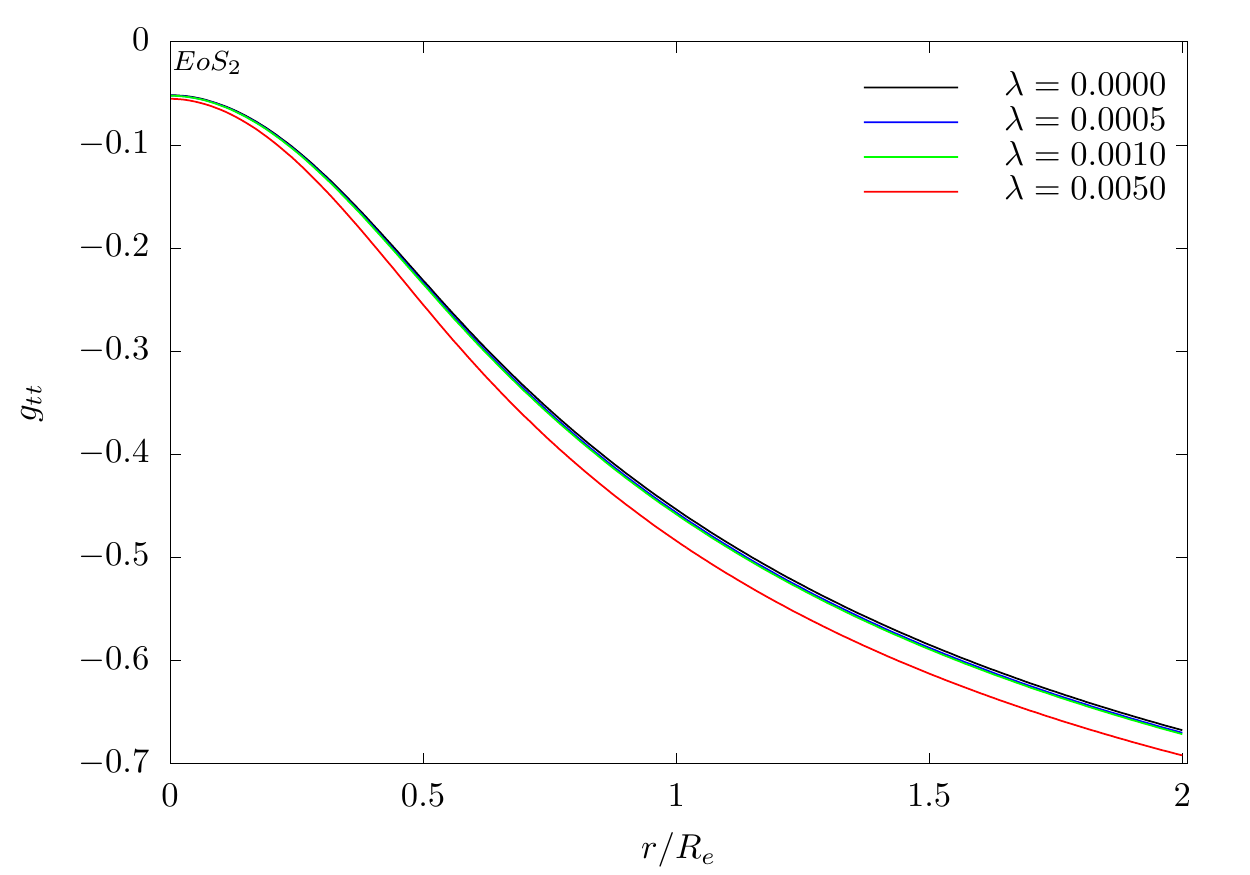}
     \end{subfigure}
        \caption{The component $g_{tt}$ of the metric tensor as a function of the ratio $R/R_e$ at the equator $(\theta=\pi/2)$, for stars rotating at the mass-shedding limit with a fixed central energy density. The curves for $\lambda=0$ correspond to the GR case.}
        \label{fig5}
\end{figure}

Lastly we analyzed in Figure \ref{fig5} the effect of Rastall's modification on the $g_{tt}$ term of the metric for stars rotating at the mass shedding limit $\Omega_{K}$. In this plots the values of $g_{tt}$ are negative due to the metric signature. In the left panel, all four curves are for stars with central energy density of $\epsilon_{c}/c^{2}=2.3\times 10^{15}g/cm^3$, and on the right side all curves have $\epsilon_{c}/c^{2}=2.4\times 10^{15}g/cm^3$. We can observe that in both EoSs the effect of increasing the parameter $\lambda$ is to decrease the deformation in space-time, but in the same way as in the other quantities analysed, the effect of Rastall's is smaller for the second EoS. 
 
\section{Conclusions}

In this work we have studied rapidly rotating compact stars in the Rastall modified theory of gravity. We solved numerically the field equations obtained using the KEH method, by applying the necessary modifications. We employed a polytropic EoS which two different choices of parameters that are widely used in the literature, the first EoS referred as $EoS_1$ can be considered a soft EoS and has maximum mass around $1.4M_{\odot}$ in the static case, the second one, $EoS_2$, that can be considered a representative of a more stiff EoS, has maximum mass above $2M_{\odot}$ in the non-rotating case. Three values for the constant $\lambda$ were implemented, $\lambda=5\times10^{-4}$, $1\times10^{-3}$ and $5\times10^{-3}$, beyond the value $\lambda=0$ that correspond to GR case, and it was assumed an uniform rotation. 

The properties of the solutions found were calculated and examined for the situations of interest. In all cases the analysed properties of stars with $EoS_2$ were less affected by the increasing of the Rastall's parameter than the ones with $EoS_1$.

We have investigated the mass-radius relation associated to rapidly rotating compact stars and static stars. In particular, our results for the static case coincided with the ones found in literature. In the case of stars rotating at the Kepler limit, we observed that the maximum mass slightly decreases while in the non-rotating case it slightly increases, and in both situations the radius of the stars is increased.

We also have explored the relations of the mass and the angular velocity at the Keplerian limit with the central energy density, and we concluded that as we increase the Rastall's parameter both the mass and the angular velocity at the Keplerian limit decrease for stars with the same central energy density. Moreover, we examined the effect of Rastall's theory on the relation between the moment of inertia and the mass, and we could see that in general the effect of increasing the parameter $\lambda$ is shifting the $I\times M$ curves down and left. However, in some cases for the Rastall theory, there is a deviation in this behavior and the moment of inertia begins to increase as the mass decreases, we believe this behavior occurs because for stars with small masses the effect of Rastall's gravity to increase the radius can be quite intense.

Our last analysis was about how the component $g_{tt}$ of the metric tensor changes due do Rastall's gravity, we can infer that the behavior of this quantity is a consequence of the previous results. Since the effect of increasing the parameter $\lambda$ is to produce stars that have smaller masses, larger radius and that rotate more slowly, it is expected that the deformation in space-time is smaller, as we can see in Figure \ref{fig5}.  

Since this is just a first study of the effects of Rastall's gravity on rapidly rotating compact stars, in the future it would be interesting to amplify the investigation of rotating stars in this modified theory of gravity. One way to do that is by applying more realistic EoSs, another possibility is analysing the scenario of differential rotation, a study of the I-Love Q relations in the context of Rastall's gravity would also be pertinent.

\section{Acknowledgements}

We thank Giovanni Formighieri  for discussions and, F.M.S. would like to thank the Coordenação de Aperfeiçoamento de
Pessoal de Nível Superior (CAPES) for financial support. L. C. N. Santos would like to thank Conselho Nacional de Desenvolvimento Científico e Tecnológico (CNPq) for partial financial support through the research Project No. 164762/2020-5.

\bibliographystyle{ieeetr}
\bibliography{ref.bib}

\end{document}